\documentclass[twocolumn,showkeys,preprintnumbers,amsmath,amssymb]{revtex4}
\usepackage{graphicx}% Include figure files
\usepackage{dcolumn}% Align table columns on decimal point
\usepackage{bm}% bold math
\begin{document}

%\preprint{APS/123-QED}

\title{Exploring hints for dark energy density evolution in light of recent data}

\author{V\'ictor H. C\'ardenas}
\affiliation{%
Instituto de F\'{\i}sica y Astronom\'ia, Facultad de Ciencias,
Universidad de Valpara\'iso, Gran Breta\~na 1111,
Valpara\'iso, Chile and \\
Centro de Astrof\'isica de Valpara\'iso, Gran Breta\~na 1111,
Valpara\'iso, Chile
}%


\begin{abstract}
Considering a quadratic parametrization of the dark energy density,
we explore signatures of evolution using data from gas mass fraction
in clusters, type Ia supernova, BAO and CMB. We find -- excluding
CMB data -- a preference for a evolution of $\rho_{de}(z)$ towards
smaller values as redshift increases, a result consistent with a
recent study using the BAO DR11 data by Delubac et al. (2015).
\end{abstract}

%Use showkeys class option if keyword
\maketitle

\section{Introduction}

The $\Lambda$CDM model is the simplest cosmological model that fits
a varied set of observational data; type Ia supernova (SNIa), baryon
acoustic oscillations (BAO), Cosmic microwave background radiation
(CMBR), growth of structure etc \cite{pero}. In this setup the
cosmological constant $\Lambda$ drives the current accelerated
expansion of the universe, detected for the first time using type Ia
supernovae \cite{snia1, snia2}. Although successful in fitting the
data, the model is awkward in many ways: for example, we do not know
the mechanism to produce such a constant in the first place. We also
do not expect to live in a special epoch where the contribution of
this constant is of the same order of magnitude than the non
relativistic matter contribution. This problem in particular is
known as the ``cosmic'' coincidence problem.
%ok

From a theoretical point of view, it is most natural to think that
this contribution comes from an evolving source (with epoch) whose
connection with the universe expansion is under study. Dark energy
(DE) is the name of this mysterious source \cite{dereview}.
%ok

Different DE models have been proposed to provide the mechanism that
explains the observational data. There are models where a new field
component is assumed to fill the universe, known as quintessence
\cite{quinta1,quinta2,quinta3,quinta4,quinta5,quinta6}, and models
where the mechanism is triggered by using a modified gravity theory
\cite{Tsujikawa2010, Capozziello2011, Starkman2011}.
%ok

In the absence of consensus regarding a theoretical description for
cosmic acceleration, theorists have proposed using the equation of
state (EoS) parameter $w(a)=p/\rho$, where $a$ is the scale factor,
as a useful phenomenological description \cite{dereview}.
%ok

In this context in \cite{shafi2009}, using the Constitution data set
for SNIa \cite{constitution}, and the Chevalier-Polarski-Linder
(CPL) parameterization for $w(a)$ \cite{Chevallier:2000qy,
Linder:2002et},
\begin{equation}\label{cpl}
 w(a)=w_0+(1-a)w_1,
\end{equation}
with $w_0$ and $w_1$ being free parameters to be fixed by
observations, the authors found a reconstructed deceleration
parameter that apparently shows a rapid variation at small redshift
around $z\simeq 0.2$. However, once the baryon acoustic oscillations
(BAO) and cosmic microwave background (CMB) data are added into the
analysis, the best fit result changes completely, showing no sign of
variation at the small redshift in agreement with what is expected
in the $\Lambda$CDM model. In \cite{Li:2010da} similar results were
found, under the assumption of a flat universe using the Union 2
data set \cite{Union2}. In \cite{Cardenas:2011a} we revisit this
problem using the Union 2 data set extending the analysis to allow
for curved spacetime.
%ok

In \cite{Cardenas:2013roa}, using data from gas mass fraction in
galaxy clusters $f_{gas}$, we encountered the same apparent behavior
found previously using SNIa \cite{shafi2009, Li:2010da,
Cardenas:2011a}.
%ok

SNIa are standardizable candles from which we measure the luminosity
distance. In the case of the gas mass fraction, we measure the X-ray
emission, which enable us to estimate the baryonic (mostly gas) and
total mass, assuming the intracluster gas is in hydrostatic
equilibrium, from which we measure the angular diameter distance to
the cluster \cite{Sasaki:1996zz}. Because the $f_{gas}$ data span a
similar redshift range as the SNIa, but depends on a completely
different physics, this finding is certainly intriguing.
%ok

Although the statistical significance of this effect is small, the
consistency between the results using SNIa and $f_{gas}$, moves us
to deepen the study of this effect at low redshift.
%ok

We also studied the possible dependence of this result - a low
redshift transition of the deceleration parameter - with different
parameterizations. In \cite{juan} we used five different types of
parameterizations and the result was always consistent with that
found using CPL. However, the analysis based on using $w(z)$
increases the errors in the parameters we want to constrain. The
problem with using $w(z)$ as the focus of study was demonstrated in
\cite{Maor:2000jy} (see also \cite{Maor:2001ku}). The essential
problem is the observational quantity, as the luminosity distance or
the angular diameter distance, depends on $w(z)$ through a double
integral smearing out the information about $w(z)$ itself and its
time variation.
%ok

As the $\Lambda$CDM model is by definition a model with a constant
DE density, in this work we focus on signals of a possible departure
from this trend. In this context, as was explained in the previous
paragraph, is not efficient to use $w(z)$ or a particular
parametrization of it; instead, we work directly with the dark
energy density, whatever that may be. This strategy was started in
\cite{Wang:2001ht}, and \cite{Wang:2001da}, where the authors
demonstrated the advantage of using the energy density instead of
the EoS parameter as the main probe to constraint.
%ok

In this paper we investigate the possibility of evolution of the
dark energy density in light of recent data. We use gas mass
fraction in clusters \cite{Sasaki:1996zz} - 42 measurements of
$f_{gas}$ in clusters extracted from \cite{Allen:2007ue} - and also
type Ia supernovae (SNIa) from the Lick Observatory Supernova Search
(LOSS) compilation sample \cite{Ganeshalingam:2013mia}. We also
consider the constraints obtained from BAO and CMB. The BAO
measurements considered in our analysis are obtained from the
WiggleZ experiment \cite{2011MNRAS.tmp.1598B}, the SDSS DR7 BAO
distance measurements \cite{2010MNRAS.401.2148P}, and 6dFGS BAO data
\cite{2011MNRAS.416.3017B}. We also include background CMB
information by using the Planck data \cite{ade:2014} to probe the
expansion history up to the last scattering surface.
% Why do not use the Planck data?
We have also perform the analysis using the WMAP 9-yr covariance
matrix from \cite{Wang:2013mha}, with no significant changes.
%ok

The paper is organized as follows: in the next section we describe
what we have learned from the $w(z)$ parametrization. Then, we
describe how to implement the interpolation method to constrain the
DE density model using the observational data available. After that,
we present the results of our study, first using SNIa and $f_{gas}$
data and then within a joint analysis. We end with a discussion of
the results.
%ok

\section{Insights from the reconstructed deceleration parameter}

Observational cosmology is essentially based on quantities derived
from the Hubble function. For example, using both type Ia supernova
or galaxy cluster data, the key functions are written in terms of
the comoving distance from the observer to the redshift $z$ given by
\begin{equation}\label{comdistance}
r(z) =  \frac{c}{H_0} \frac{1}{\sqrt{-\Omega_k}} \sin
\sqrt{-\Omega_k} \int_0^z \frac{dz'}{E(z')},
\end{equation}
where $E(z)=H(z)/H_0$ contains the cosmology. For example, for the
case of the $\Lambda$CDM model the function is,
\begin{equation}\label{edez}
E^2(z)  = \Omega_m (1+z)^3+\Omega_r (1+z)^4+ \Omega_k (1+z)^2 +
\Omega_{\Lambda}.
\end{equation}
Here $\Omega_{m}$ comprise both the baryonic and non baryonic DM. We
know the radiation component is negligible at low redshift; in fact,
we know $ h^2 \Omega_r = 2.47 \times 10^{-5}$ from \cite{ade:2014}.
However, if we want to constrain our model using data from BAO and
CMB, we have to use it, because these probes refers to both the last
scattering redshift and the drag epoch.
%ok

In practice, by using the CPL parameterization (\ref{cpl}) for the
DE component, and after testing it against the observational data,
we get the best fit values of the parameters, which give us the best
Hubble function $E (z) \equiv H (z) /H_0$ that agrees with the data.
From it, following previous works \cite{shafi2009, Li:2010da,
Cardenas:2011a}, we reconstruct the deceleration parameter function
\begin{equation}\label{qdzeq}
q(z) = (1+z)\frac{1}{E(z)}\frac{dE(z)}{dz}-1.
\end{equation}
In order to motivate the next section, we will repeat the
calculation with recent data. We use gas mass fraction in clusters
extracted from \cite{Allen:2007ue}, and also type Ia supernovae
(SNIa) from the LOSS compilation sample
\cite{Ganeshalingam:2013mia}. From now on we assume a spatially flat
universe ($\Omega_k=0$).
%ok

The SNIa data give the luminosity distance $d_L(z)=(1+z)r(z)$. We
fit the SNIa with the cosmological model by minimizing the $\chi^2$
value defined by
\begin{equation}
\chi_{SNIa}^2=\sum_{i=1}^{586}\frac{[\mu(z_i)-\mu_{obs}(z_i)]^2}{\sigma_{\mu
i}^2},
\end{equation}
where  $\mu(z)\equiv 5\log_{10}[d_L(z)/\texttt{Mpc}]+25$ is the
theoretical value of the distance modulus, $\mu_{obs}$ is the
corresponding observed one, and $\sigma_{\mu i}$ is the error
associated to it. As explained in \cite{Ganeshalingam:2013mia}, the
error comprises three components: the uncertainty from light-curve
fits, a component due to the peculiar velocity of each SNIa, and an
intrinsic scatter term which depends on the sample (see Table 1 in
\cite{Ganeshalingam:2013mia}).
%ok

The gas mass fraction data we use span a redshift range
$0.05<z<1.1$. The $f_{gas}$ data are quoted for a flat $\Lambda$CDM
reference cosmology with $h=H_0/100$ km s$^{-1}$Mpc$^{-1}=0.7$ and
$\Omega_M=0.3$. To obtain the restrictions we use the model function
from \cite{Allen:2004cd}:
\begin{equation}\label{fgas}
f_{gas}^{\Lambda CDM}(z)=\frac{b \Omega_b}{(1+0.19\sqrt{h})
\Omega_M} \left[\frac{d_A^{\Lambda CDM}(z)}{d_A(z)} \right]^{3/2},
\end{equation}
where $b$ is a bias factor motivated by gas-dynamical simulations
which suggest the the baryon fraction in clusters is slightly lower
than for the universe as a whole. From \cite{eke98} $b=0.824 \pm
0.0033$ is obtained. Following \cite{Allen:2004cd} we adopt a
gaussian prior on $b$, taking into account systematic uncertainties,
so we use $b=0.824 \pm 0.089$. In the analysis we also use standard
Gaussian priors on $\Omega_b h^2 = 0.02205 \pm 0.00028$ and $h=0.72
\pm 0.08$ from Planck and WMAP polarization \cite{ade:2014}.
%ok

The use of SNIa and $f_{gas}$ data separately, as demonstrated in
\cite{Cardenas:2013roa}, generates a behavior that is consistent
between them. For that reason, in what follows we show first the
result considering both probes together. Given the two data sets are
consistent each other, we use the standard $\chi^2$ analysis.
%ok

In the analysis (see the details in Appendix A) we consider
$h,\Omega_m, w_0, w_1,\Omega_b$ and $b$ as free parameters. As we
mentioned, we have added Gaussian priors for $h, \Omega_b$ and $b$.
After the analysis the best fit values are those shown in Table
\ref{tab:table00}.

\begin{table}
\caption{\label{tab:table00} The best fit values for the free
parameters using SNIa + $f_{gas}$. See also Fig. \ref{fig:figure3}.}
\begin{tabular}{ccc}
%Left\footnote{Note a.}&Centered\footnote{Note b.}&Right\\
\hline
$h$ & $\Omega_m$ & $w_0$ \\
\hline
0.695$\pm$0.004 & 0.30$\pm$0.04 & -0.73$\pm$0.16 \\
\hline
$w_1$ & $\Omega_b$ & $b$ \\
\hline -2.7$\pm$1.5 & 0.0457$\pm$0.0008 & 0.84$\pm$0.09
\end{tabular}
\end{table}

\begin{figure}
  \begin{center}
    \includegraphics[width=7cm]{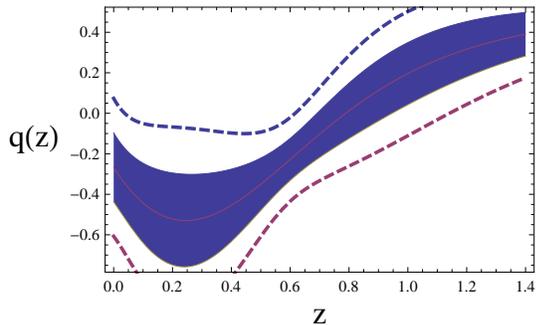}
  \end{center}
\caption{ Using the LOSS compiled sample by
\cite{Ganeshalingam:2013mia} and the $ f_{gas} $ data from
\cite{Allen:2004cd} we plot the reconstructed deceleration parameter
(\ref{qdzeq}) using the best fit values for the CPL parametrization.
We consider the error propagation at one and two sigmas in the best
fit parameters. We observe a hint for a low redshift transition,
reaching the maximum acceleration around $ z \simeq 0.2 $, and later
a slowing down of the acceleration. The shaded region is $1\sigma$
and the region between the dashed lines is $2\sigma$.}
\label{fig:figure3}
\end{figure}

Using the best fit values for the CPL parameters ($w_0,w_1$), the
deceleration parameter (\ref{qdzeq}), with error propagation, is
shown in Fig.\ref{fig:figure3}.
%ok
From figure \ref{fig:figure3}, we notice that the combined action of
SNIa and $ f_ {gas} $ data suggest a universe in transit, from a
decelerated expansion regime to an accelerated one, with the
transition redshift $ z \simeq 0.8 $, in agreement with
$\Lambda$CDM, and also a slowing down of the acceleration at recent
times, a result that seems to be supported at a $2\sigma$ level.
%ok

It is important to stress here that the support for this low
redshift behavior disappears once we consider BAO and CMB data. This
result gives us a hint that seems to indicate a tension between low
and high redshift probes.
%ok

The key result here, which as far as we know no one has mentioned to
date, is that in all the previous cases studied
\cite{Cardenas:2011a,Li:2010da,Cardenas:2013roa} showing a
reconstructed deceleration parameter with a rapid variation at the
small redshift, i.e. using SNIa and $f_{gas}$ data only, the
reconstructed DE density appears to be a decreasing function with
increasing redshift at $2\sigma$.
%ok

In fact, in the special case of the CPL parametrization for $w(z)$
Eq.(\ref{cpl}), we find that
\begin{equation}\label{cplX}
X(z) \equiv \frac{\rho_{de}(z)}{\rho_{de}(0)}= e^{-\frac{3 w_1
z}{1+z}} (1+z)^{3 (1 + w_0 + w_1)},
\end{equation}
which can be interpreted as a very special parametrization for the
DE density. The best fit values of the parameters enable us to get
the Hubble function $E(z)$ and the DE density from (\ref{cplX}). We
have checked that the same trend (at $2\sigma$ level) is obtained by
using both the Constitution data set and the Union 2 set together
with the gas mass fraction data \cite{Allen:2004cd}.
%ok

\begin{figure}
  \begin{center}
    \includegraphics[width=7cm]{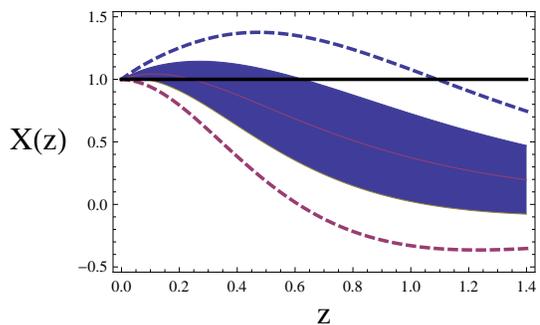}
  \end{center}
  \caption{Using the SNIa and the gas mass fraction data used in the previous analysis
  we plot the DE density (\ref{cplX}) reconstructed using the best fit values for the
CPL parametrization. We consider the error propagation at one and
two sigmas in the best fit parameters. We observe a preference for a
decaying DE density with increasing redshift.}
  \label{fig:figure4}
\end{figure}

Here, using recent SNIa data and $f_{gas}$ data, we have
reconstructed the DE density from (\ref{cplX}) and displayed it in
Fig.\ref{fig:figure4}. The data through the CPL parametrization
seems to suggest an evolving DE density at a $2\sigma$ level.
%ok

%%%%%%%%%%%%%%%%%%%%%%%%%%%%%%%%%%%%%%%%

In what follows we test whether the behavior that seems to suggest
the low-redshift data -- a decreasing DE density with increasing
redshift using a parametrization for $w(z)$ -- persist, once we
consider a parametrization for the DE density.
%ok

\section{The Method}

In order to explore the eventual redshift evolution of the DE
density, we use as a probe the definition
$X(z)=\rho_{de}(z)/\rho_{de}(0)$, and write the Hubble function as
\begin{equation}\label{edez2}
E^2(z)  = \Omega_m (1+z)^3+\Omega_r (1+z)^4 + \Omega_X X(z),
\end{equation}
where $\Omega_r = 2.47 \times 10^{-5} h^{-2}$ and  $\Omega_m +
\Omega_r + \Omega_X=1$. In the case of using the CPL parametrization
Eq.(\ref{cpl}), we have already found the expression (\ref{cplX}).
%ok

In this work we use the method suggested by \cite{Wang:2001ht},
\cite{Wang:2001da}, and extended by \cite{Wang:2003cs},
parameterizing the DE density through a quadratic interpolation with
two free parameters. In this work we restrict ourselves to this
number of free parameters, just to compare with previous ones
\citep{Wang:2004ru} and maintain a meaningful statistical analysis.
In assuming a quadratic function for $X(z)$, it is convenient to
define the free parameters in reference to the maximum redshift
value in the data. In this case we use
\begin{equation}\label{inter1}
X(z)= 1+\frac{z (4f_1 -f_2-3)}{z_m}-\frac{2 z^2 (2
   f_1-f_2-1)}{z_m^2},
\end{equation}
where $z_m$ is the maximum redshift value in the data, and the free
parameters are: $f_1=X(z_m/2)$ and $f_2=X(z_m)$. If there is no
evolution, i.e. a cosmological constant is preferred, both
parameters should have to be equal to unity.
%ok

\section{Results}

To start the analysis, we test the parameterized DE density
(\ref{inter1}) for each of the observational probes separately. The
idea here is to clarify what kind of trend suggests the use of each
of the different data sets considered in this work, and also
evaluate the consistency among them. In what follows we have used
$z_m=1.34$, which is the highest redshift in our data. We have tried
different values for $z_m$ and find no variation in the qualitative
behavior.
%ok

To study the consistency among the different data sets with the
$\Lambda$CDM model, we follow the method proposed in
\cite{perivola2010}, where a distance $d_{\sigma}$ (in units of
$\sigma$) from the best fit point to the $\Lambda$CDM model, is
defined through the relation
\begin{equation}\label{dsigma}
1 - \Gamma(1,\Delta \chi^2/2)/\Gamma(1) =
\verb"Erf"(d_{\sigma}/\sqrt{2}),
\end{equation}
where the left hand side is the cumulative distribution function
(for two parameters), and $\Delta \chi^2 =
\chi^2_{(f_1,f_2)}-\chi^2_{min}$ is the $\chi^2$ difference between
the best fit and the $\Lambda$CDM point ($f_1=f_2=1$).
%ok

Using the LOSS compiled sample \cite{Ganeshalingam:2013mia} of SNIa,
assuming a flat universe ($\Omega_k=0$) with the free parameters
being, $\Omega_m$, $h$, $f_1$, $f_2$, we obtain a distance
$d_{\sigma}=2.46
 \sigma$ away from the reference point ($f_1=f_2=1$): the
$\Lambda$CDM model. Using the gas mass fraction $f_{gas}$ data from
\cite{Allen:2007ue}, with the free parameters being $\Omega_m$, $h$,
$f_1$, $f_2$, $\Omega_b$ and $b$, and using the same priors on $b$
and $\Omega_b$ mentioned in section II, we obtain a distance
$d_{\sigma}=1.896 \sigma$ away from the reference point
($f_1=f_2=1$). Using the BAO data, with $\Omega_m$, $h$, $f_1$,
$f_2$, and $\Omega_b$ as free parameters, we obtain a distance
$d_{\sigma}=0.976 \sigma$ away from the reference point
($f_1=f_2=1$). Using the CMB data, with the same free parameters as
in the BAO set, we get $d_{\sigma}= 1.107 \sigma$ away from the
$\Lambda$CDM model. A summary of our results are shown in table
\ref{tab:table10}.
%ok

\begin{figure}
  \begin{center}
    \includegraphics[width=7cm]{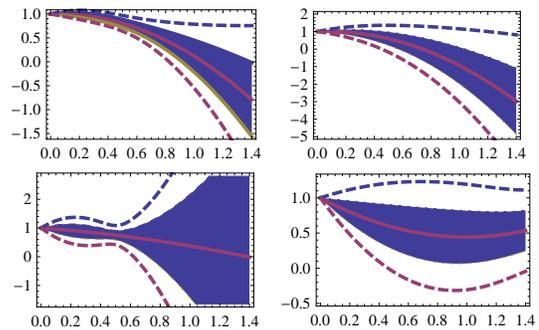}
  \end{center}
 \caption{The horizontal axes indicate the redshift $z$ and the vertical axes indicate $X(z)$.
The upper left graph is reconstructed using SNIa data, the upper
right graph was obtained using $f_{gas}$ data, the lower left graph
is based on BAO data, and the lower right is obtained using CMB
data. All the graphs show $X(z)$ as a function of redshift, with
error propagation at one (shaded area) and two (dahed lines) sigmas.
Notice the trend of a decaying DE density with redshift in the case
of SNIa and $f_{gas}$.}
  \label{fig:figure0}
\end{figure}

Notice the best fit values of $\Omega_m$ for each data set. The DE
evolution model we are testing prefer rather lower values for
$\Omega_m$ ($\simeq 0.26$) once we test it against BAO and CMB data.
This fact makes these data sets consistent each other. Meanwhile,
using SNIa and $f_{gas}$, the DE evolution model takes values
$\Omega_m \simeq 0.3$.
\begin{table}
\caption{\label{tab:table10} For each one of the data sets we
display the distance from the best fit point (see the definition in
Eq. (\ref{dsigma})) in the two-dimensional space $(f_1,f_2)$, from
the $\Lambda$CDM model. See also Fig. \ref{fig:figure0}.}
\begin{tabular}{ccccc}
%Left\footnote{Note a.}&Centered\footnote{Note b.}&Right\\
\hline
Set & $d_{\sigma}$ & $\Omega_m$ & $f_1$ & $f_2$ \\
\hline
SNIa & 2.46 & 0.299 & 0.627 & -0.635 \\
$f_{gas}$ & 1.896 & 0.293 & 0.196 & -2.622 \\
CMB & 1.107 & 0.256 & 0.503 & 0.509 \\
BAO & 0.976 & 0.262 & 0.637 & 0.049

\end{tabular}
\end{table}
A reconstruction of the DE density for each one of the data sets is
plotted in Fig. \ref{fig:figure0}.
On the one hand, from the figure it is clear the similarity in the
result between SNIa and $f_{gas}$. Under general considerations
(neglecting an initial mild increase in $X(z)$) we obtain a
reconstructed DE density evolving, apparently decreasing as the
redshift increases. In this sense, these two probes are consistent
each other. On the other hand, both the results using with BAO and
CMB are essentially consistent with a cosmological constant even at
one sigma, thus being inconsistent with the previous two probes.
%ok

This result is actually what we have taken into account in the
analysis performed in \cite{Cardenas:2013roa}, where we separate the
analysis using first SNIa + $f_{gas}$ and then all the probes
together. In what follows, we adopt the same procedure starting with
a joint analysis between SNIa + $f_{gas}$, and after that a joint
analysis with all the data set together.
%ok

First, using the LOSS compiled sample \cite{Ganeshalingam:2013mia}
and the gas mass fraction $f_{gas}$ data from \cite{Allen:2007ue},
we obtain the best fit values of the model (\ref{inter1}), assuming
a flat universe ($\Omega_k=0$) with the free parameters being,
$\Omega_m$, $h$, $f_1$, $f_2$, $\Omega_b$ and $b$.
%ok

\begin{figure}
  \begin{center}
    \includegraphics[width=7cm]{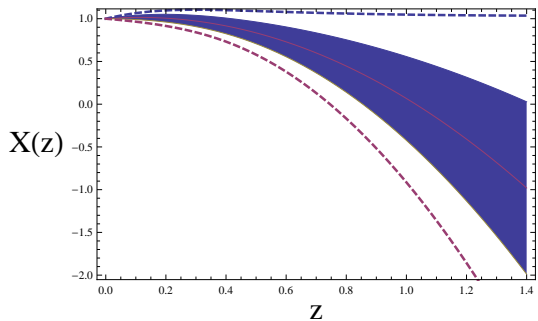}
  \end{center}
 \caption{Adopting the best fit parameters obtained from using the LOSS compilation by \cite{Ganeshalingam:2013mia}
of SNIa together with the gas mass fraction data from
\cite{Allen:2004cd} (see Table \ref{tab:table01}), we plot the DE
density as a function of redshift, with error propagation at one
(shaded area) and two (dahed lines) sigmas, including error
correlation between the parameters $f_1$ and $f_2$. Notice the
intriguing result of a decaying DE density with redshift.}
  \label{fig:figure1}
\end{figure}

Using these data together, using the same priors on $b$ and
$\Omega_b$ mentioned in section II, we find the best fit values
shown in Table \ref{tab:table01}. We plot the DE density as a
function of redshift in Fig.(\ref{fig:figure1}), with error
propagation at one and two sigmas. Here we have taken into account
the correlation error between the parameters $f_1$ and $f_2$. Note
that at one and two sigmas, the reconstructed DE density appears to
decrease with increasing redshift.

\begin{table}
\caption{\label{tab:table01} The best fit values for the free
parameters using (1) SNIa + $f_{gas}$, and (2) SNIa +
$f_{gas}$+BAO+CMB. See also Figs. \ref{fig:figure1}, and
\ref{fig:figure2}.}
\begin{tabular}{cccc}
%Left\footnote{Note a.}&Centered\footnote{Note b.}&Right\\
\hline
Set & $h$ & $\Omega_m$ & $f_1$ \\
\hline
(1) & 0.698$\pm$0.004 & 0. 29$\pm$0.03 & 0.65$\pm$0.22\\
(2) & 0.704$\pm$0.004 & 0.273$\pm$0.009 & 0.87$\pm$0.10 \\
\hline
Set & $f_2$ & $\Omega_b$ & $b$ \\
\hline
(1) & -0.75$\pm$0.93 & 0.0453 $\pm$0.0007 & 0.83$\pm$0.06 \\
(2) & 0.89$\pm$0.64 & 0.0457$\pm$0.0008 & 0.79$\pm$0.03

\end{tabular}
\end{table}

% Negative DE density
Obtaining a decreasing DE density as the redshift increases, as the
one obtained here, would eventually lead us to get negative values
for $X(z)$, as can be observed in Fig. \ref{fig:figure1} for
$z>0.8$. Although this idea may seem contrary to common sense, a
negative DE density has been considered in the past. For example, in
\cite{Ahmed:2002mj} the author considered a model inspired from
unimodular gravity, predicting fluctuations in the cosmological
constant. These fluctuations of the cosmological ``constant'' are
always of the order of the ambient density, and it is not strange
that $\Lambda$ may eventually take negative values. Also in this
context and using the back-reaction approach the author of
\cite{Brandenberger:2002sk} obtains a cosmological constant which
oscillates about $1/2$ the total $\Omega$ on Hubble time scales. It
is also interesting to mention the work done in
\cite{Quartin:2008px}. There the authors performed a study of an
interacting DM/DE model using a moderately general interaction term.
One of their conclusions was that, based on their examples, a
solution to the coincidence problem would require that the DE
density should take negative values in the past.
%ok

Once we consider the data from BAO and CMB together with the already
described SNIa and $f_{gas}$, we minimize the joint chi square,
% explain the method of the weighted chi-squared
$\chi^2=\chi^2_{SNIa}+\chi^2_{BAO}+\chi^2_{CMB}+\chi^2_{fgas}$,
where each element is defined in appendix A. In the conventional
joint $\chi^2$ analysis, we weight each probe equally. This may be
problematic if two data sets are mutually inconsistent
\cite{Lahav:1999hu}. A well-motivated approach to assigning weights
is the ``hyper-parameter'' method \cite{Lahav:1999hu,Hobson:2002zf}.
In this approach, finding the best-fitting parameters requires us to
minimize an effective $\chi^2$ given by
\begin{equation}\label{efchi2}
\chi^2_{hy}=\sum_i N_i \ln \chi^2_i,
\end{equation}
where $i$ sums over all the data sets ($i=$ SNIa, BAO, CMB, fgas),
and $N_i$ is the number of data points in each data set. Once the
$\chi^2$ values have been obtained, we find the posterior
distribution for the parameters using the conventional $\chi^2$ or
the hyper-parameter version $\chi^2_{hy}$ \citep{Hobson:2002zf}. The
results are shown in Table \ref{tab:table01}.

\begin{figure}
  \begin{center}
    \includegraphics[width=7cm]{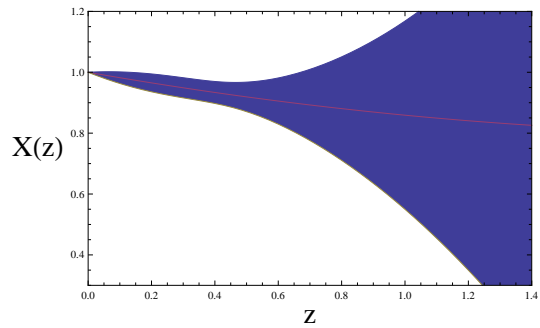}
  \end{center}
 \caption{Adopting the best fit parameters from the analysis using SNIa + $ f_ {gas} $ +
BAO + CMB, we plot the DE density as a function of redshift, with
error propagation to one sigma including error correlation between
the parameters $f_1$ and $f_2$. The result shows that the addition
of these probes makes the reconstructed DE density $X(z)$ consistent
with no evolution.}
  \label{fig:figure2}
\end{figure}

A quadratic parametrization of this type was used by
\cite{Wang:2004ru}, where the authors used a sample with 192 SNIa in
combination with CMB and LSS data. Unfortunately, the authors did
not show the result using only SNIa data. When comparing their
results with our analysis, considering SNIa, $f_{gas}$ and both BAO
and CMB data, our results show smaller uncertainties and are more
consistent with no change compared with them. In fact, their results
indicate a growing DE density with redshift at one sigma. We plot
the DE density as a function of redshift in Fig.(\ref{fig:figure2}),
with error propagation at one sigma.
%%%%%%%%%%%%%%%%%%%%%%
It is very clear that the addition of BAO and CMB data makes the
evidence of evolution disappears. This is exactly what we discussed
at the end of Section II, in the context of a CPL parameterization
for $w(z)$.
%ok
% Link the actual result with the previous one.
At this point it seems there is an intriguing connection between the
previously found low redshift transition of the deceleration
parameter $q(z)$, and a DE density  $X(z)$ that decreases as the
redshift increases.
%ok

It is also interesting to mention that such a behavior -- a low
redshift transition of the deceleration parameter $q(z)$
(\ref{qdzeq}) -- was previously found first (as far as we know) by
\cite{Vanderveld:2006rb} in the context of Lemaitre-Tolman-Bondi
inhomogeneous models. In that work, and in recent ones
\cite{february2010, Bengochea:2014iha}, the authors derived an
effective deceleration parameter for void models, indicating that
such a behavior of $q(z)$ may be considered a signature for the
existence of voids.
%ok

Furthermore, a decreasing DE density with increasing redshift is
also found in the recent BAO data release by \cite{Delubac:2014aqe},
where a tension between BAO data and CMB is found. This tension
reveals that in order to accommodate these new data it is not
sufficient to go into models with non-zero curvature or a constant
$w \neq -1$ DE, essentially because the data requires a decreasing
$D_A(z=2.34)$ while increasing $D_H(z=2.34)$. The intriguing result,
assuming a flat universe with dark matter and DE is that (quoted
from \cite{Delubac:2014aqe})
\begin{equation}
\frac{\rho_{de}(z=2.34)}{\rho_{de}(z=0)}=-1.2 \pm 0.8,
\end{equation}
which shows this data seems to favor an evolving DE density as we
have found in this work using SNIa and $f_{gas}$.
%ok

An evolving DE density, as seems to be suggested by our study, not
only means that the DE contribution itself varies with time. This
may also be produced by other means. For example, it can be produced
by a locally inhomogeneous distribution of matter, like the
previously mentioned example using LTB models. Also, as we have
mentioned on our discussion on negative DE density, this can also be
obtained by assuming an explicit interaction between DM and DE (see
for example \cite{Quartin:2008px}).
%ok

After having uploaded a draft version of this paper to the web
\cite{Cardenas:2014jya}, several works have appeared whose results
point in the same direction. In \cite{Sahni:2014ooa} the authors use
the BAO data from the BOSS DR11 \cite{Delubac:2014aqe}, along $H(z)$
measurements at low redshift, finding a considerable tension with
the standard $\Lambda$CDM model, implying a evolution for DE. Also
in \cite{Abdalla:2014cla}, the authors use the same data from BOSS
to show that such a departure from the $\Lambda$CDM model can be
accommodated assuming an interaction between dark matter and DE,
excluding the null interaction at 2$\sigma$.
%ok

%%%%%%%%%%%%%%%%%%%%%%%%%%%%%%%%%%%%%%%%%%%%%%%%%%%%%%%%%%%%%%%%%%%
\section{Discussion}

In this paper we have presented a study of possible signs of
evolution of the DE density in light of recent data. We use gas mass
fraction in clusters - 42 measurements of $f_{gas}$ in clusters
extracted from \cite{Allen:2007ue} - and also type Ia supernovae
data compiled in \cite{Ganeshalingam:2013mia} from the LOSS team. We
also consider the constraints obtained by adding measurements from
baryon acoustic oscillations (BAO) and cosmic microwave background
radiation (CMB). We have found -- using SNIa and $f_{gas}$ data --
evidence that relates the previously found low redshift transition
of the deceleration parameter to a decreasing DE density evolution
with increasing redshift. This result seems to confirm the tension
between the data at low redshift and those from CMB. This result is
also consistent with a recent anisotropic BAO measurement of BOSS
DR11 \cite{Delubac:2014aqe}, which shows that the data appear to
prefer a decreasing DE density with increasing redshift.
%ok

Although the statistical significance of the result is low -- this
manifests up to $2\sigma$ -- it is interesting to focus on what the
low redshift data are telling us. Because we expect the DE component
to be dominant at recent (low redshift) epoch, and the fact that now
data from SNIa, gas mass fraction and the recent BAO DR11 results
all seem to agree on this peculiar behavior at low redshift, we may
conclude that something in our near neighborhood is producing this
result. This conclusion is also reinforced with the intriguing
similarity between our finding and the result using LTB
inhomogeneous model, where the effective deceleration parameter
shows the same transition at low redshift, assuming we live inside a
void. In summary, the analysis in this work suggests either (i) we
live inside a void, or (ii) there is an evolving DE model that
produces rapid changes at low redshift. So, it is clear that a
careful study of low redshift behavior is needed to enlighten our
understanding of DE.
%ok

\begin{acknowledgments}
The author wishes to thank Yoelsy Leyva, Juan Maga\~na and Sergio
del Campo for useful discussions. This work was funded by the
Comisi\'on Nacional de Ciencias y Tecnolog\'{\i}a through FONDECYT
Grant 1110230 and DIUV 50/2013.
\end{acknowledgments}

\appendix

\section{Statistical Analysis}

The SNIa data give the luminosity distance $d_L(z)=(1+z)r(z)$. We
fit the SNIa with the cosmological model by minimizing the $\chi^2$
value defined by
\begin{equation}
\chi_{SNIa}^2=\sum_{i=1}^{586}\frac{[\mu(z_i)-\mu_{obs}(z_i)]^2}{\sigma_{\mu
i}^2},
\end{equation}
where  $\mu(z)\equiv 5\log_{10}[d_L(z)/\texttt{Mpc}]+25$ is the
theoretical value of the distance modulus, and $\mu_{obs}$ is the
corresponding observed one.
%ok

For the analysis of the gas mass fraction, following
\cite{Allen:2004cd} using the standard priors for $\Omega_b$, $h$
and $b$ mentioned in the text, the $\chi^2$ value is
\begin{eqnarray}\label{chifgas}
\chi_{fgas}^2=\sum_{i=1}^{42}\frac{(f_{gas}^{\Lambda
CDM}(z_i)-f_{gas,i})^2}{\sigma_{fgas,i}^2} + \\
+ \frac{(\Omega_b h^2 - 0.02205)^2}{0.00028^2} +
\frac{(h-0.72)^2}{0.08^2} + \frac{(b-0.824)^2}{0.089^2}. \nonumber
\end{eqnarray}

The BAO measurements considered in our analysis are obtained from
the WiggleZ experiment \cite{2011MNRAS.tmp.1598B}, the SDSS DR7 BAO
distance measurements \cite{2010MNRAS.401.2148P} and 6dFGS BAO data
\citep{2011MNRAS.416.3017B}.

The $\chi^2$ for the WiggleZ BAO data is given by
\begin{equation}
\chi^2_{\scriptscriptstyle WiggleZ} = (\bar{A}_{obs}-\bar{A}_{th})
C_{\scriptscriptstyle WiggleZ}^{-1} (\bar{A}_{obs}-\bar{A}_{th})^T,
\end{equation}
where the data vector is $\bar{A}_{obs} = (0.474,0.442,0.424)$ for
the effective redshift $z=0.44,0.6$ and 0.73. The corresponding
theoretical value $\bar{A}_{th}$ denotes the acoustic parameter
$A(z)$ introduced by \cite{2005ApJ...633..560E}:
\begin{equation}
A(z) = \frac{D_V(z)\sqrt{\Omega_{m}H_0^2}}{cz},
\end{equation}
and the distance scale $D_V$ is defined as
\begin{equation}
D_V(z)=\frac{1}{H_0}\left[(1+z)^2D_A(z)^2\frac{cz}{E(z)}\right]^{1/3},
\end{equation}
where $D_A(z)$ is the Hubble-free angular diameter distance which
relates to the Hubble-free luminosity distance through
$D_A(z)=D_L(z)/(1+z)^2$. The inverse covariance
$C_{\scriptscriptstyle WiggleZ}^{-1}$ is given by
\begin{equation}
C_{\scriptscriptstyle WiggleZ}^{-1} = \left(
\begin{array}{ccc}
1040.3 & -807.5 & 336.8\\
-807.5 & 3720.3 & -1551.9\\
336.8 & -1551.9 & 2914.9
\end{array}\right).
\end{equation}

Similarly, for the SDSS DR7 BAO distance measurements, the $\chi^2$
can be expressed as \cite{2010MNRAS.401.2148P}
\begin{equation}
\chi^2_{\scriptscriptstyle SDSS} =
(\bar{d}_{obs}-\bar{d}_{th})C_{\scriptscriptstyle
SDSS}^{-1}(\bar{d}_{obs}-\bar{d}_{th})^T,
\end{equation}
where $\bar{d}_{obs} = (0.1905,0.1097)$ are the data points at
$z=0.2$ and $0.35$. $\bar{d}_{th}$ denotes the distance ratio
\begin{equation}
d_z = \frac{r_s(z_d)}{D_V(z)}.
\end{equation}
Here, $r_s(z)$ is the comoving sound horizon,
\begin{equation}
 r_s(z) = c \int_z^\infty \frac{c_s(z')}{H(z')}dz',
 \end{equation}
where the sound speed $c_s(z) = 1/\sqrt{3(1+\bar{R_b}/(1+z)}$, with
$\bar{R_b} = 31500 \Omega_{b}h^2(T_{CMB}/2.7\rm{K})^{-4}$ and
$T_{CMB}$ = 2.726K.

The redshift $z_d$ at the baryon drag epoch is fitted with the
formula proposed by \cite{1998ApJ...496..605E},
\begin{equation}
z_d =
\frac{1291(\Omega_{m}h^2)^{0.251}}{1+0.659(\Omega_{m}h^2)^{0.828}}[1+b_1(\Omega_b
h^2)^{b_2}],
\end{equation}
where
\begin{eqnarray}
&b_1 = 0.313(\Omega_{m}h^2)^{-0.419}[1+0.607(\Omega_{m}h^2)^{0.674}], \\
&b_2 = 0.238(\Omega_{m}h^2)^{0.223}.
\end{eqnarray}

$C_{\scriptscriptstyle SDSS}^{-1}$ in Eq. (12) is the inverse
covariance matrix for the SDSS data set given by
\begin{equation}
C_{\scriptscriptstyle SDSS}^{-1} = \left(
\begin{array}{cc}
30124 & -17227\\
-17227 & 86977
\end{array}\right).
\end{equation}

For the 6dFGS BAO data \cite{2011MNRAS.416.3017B}, there is only one
data point at $z=0.106$, the $\chi^2$ is easy to compute:
\begin{equation}
\chi^2_{\scriptscriptstyle 6dFGS} =
\left(\frac{d_z-0.336}{0.015}\right)^2.
\end{equation}

The total $\chi^2$ for all the BAO data sets thus can be written as
\begin{equation}
\chi^2_{BAO} = \chi^2_{\scriptscriptstyle WiggleZ} +
\chi^2_{\scriptscriptstyle SDSS} + \chi^2_{\scriptscriptstyle
6dFGS}.
\end{equation}

We also include background CMB information by using the Planck data
\cite{ade:2014} extracted from the analysis performed by
\cite{Wang:2013mha} to probe expansion history up to the last
scattering surface. The $\chi^2$ for the CMB data is constructed as
\begin{equation}\label{cmbchi}
 \chi^2_{CMB} = X^TC_{CMB}^{-1}X,
\end{equation}
where
\begin{equation}
 X =\left(
 \begin{array}{c}
 l_A - 301.65 \\
 R - 1.7499 \\
 z_* - 1090.41
\end{array}\right).
\end{equation}
Here $l_A$ is the ``acoustic scale'' defined as
\begin{equation}
l_A = \frac{\pi d_L(z_*)}{(1+z)r_s(z_*)},
\end{equation}
where $d_L(z)=D_L(z)/H_0$ and the redshift of decoupling $z_*$ is
given by \cite{husugi},
\begin{equation}
z_* = 1048[1+0.00124(\Omega_b h^2)^{-0.738}]
[1+g_1(\Omega_{m}h^2)^{g_2}],
\end{equation}
\begin{equation}
g_1 = \frac{0.0783(\Omega_b h^2)^{-0.238}}{1+39.5(\Omega_b
h^2)^{0.763}},
 g_2 = \frac{0.560}{1+21.1(\Omega_b h^2)^{1.81}},
\end{equation}
The ``shift parameter'' $R$ defined as \cite{BET97}
\begin{equation}
R = \frac{\sqrt{\Omega_{m}}}{c(1+z_*)} D_L(z).
\end{equation}
$C_{CMB}^{-1}$ in Eq. (\ref{cmbchi}) is the inverse covariance
matrix,
\begin{equation}
C_{CMB}^{-1} = \left(
\begin{array}{ccc}
42.722 & -419.68 & -0.7659\\
-419.68 & 57394.2 & -193.808\\
-0.7659 & -193.808 & 14.700
\end{array}\right).
\end{equation}
We have also tried the WMAP 9-yr data \cite{cmb2} finding no
significant variation in the qualitative behavior.

For all the combinations of data mentioned in the paper, we have
used the conventional joint $\chi^2$ analysis. In this case we
minimize
\begin{equation}\label{chi2mini}
\chi^2_{total} = \sum_j \chi^2_j,
\end{equation}
where each $\chi^2_j$ follows the chi-square distribution. This
procedure assumes that we trust the observational errors. When we
combine different data sets, there is the concern about the extent
to which two independent data sets are consistent with one another,
with the worst scenario being when they are completely inconsistent.
In such a case, one may wish to allow freedom in the relative
weights. The hyper-parameter approach
\cite{Lahav:1999hu,Hobson:2002zf} is a method that implements these
ideas. Essentially the method generalizes (\ref{chi2mini}) to
\begin{equation}\label{chi2hyper}
\chi^2_{total}=\sum_j \alpha_j \chi^2_j,
\end{equation}
where the $\alpha_j$ are the weight parameter for each data set.
Assuming the prior probabilities of $\log(\alpha_j)$ are uniform,
and after marginalizing over these parameters, the posterior
probability (in its gaussian form) can be written in terms of the
original $\chi^2_j$ in such a way that one should consider
minimizing
\begin{equation}\label{chi2hyper2}
\chi^2_{hyper}=\sum_j N_j \ln \chi^2_j,
\end{equation}
instead of (\ref{chi2mini}). Here $N_j$ is the number of data points
of the corresponding data set.

%\label{lastpage}

\end{document}